# ENERGY AWARE PATH SEARCH FOR SENSOR WITH PARAMETERS AS USED IN AGRICULTURAL FIELD

Smitha N Pai[1] K.C.Shet[2] Mruthyunjaya H.S.[3]

[1]Department of Computer Science and Engg. Manipal University, Manipal, India
`smitha.pai@manipal.edu`
[2]Department of Computer Engineering, Suratkal, India
`kcshet@nitk.ac.in`
[3]Department of Electronics and Communication. Manipal University, Manipal, India
`mruthyu.hs@manipal.edu`

## ABSTRACT

*Sensors placed in agricultural field should have long network life. Failure of node or link allows re-routing and establishing a new path from the source to the sink. In this paper, a new path is established such that it is energy aware during path discovery and is active for longer interval of time once it is established. The parameters used for simulation are as those used in agricultural application.*

## KEYWORDS

*Agriculture, Energy Aware, Routing, Sensor Network*

## 1. INTRODUCTION

Sensors are used to monitor environmental parameters like temperature, humidity, pressure, water content etc. Sensors gather the information and send them to the base station. Base station analyses the data and carries out further processing. In order to send the data to the base station a path is established between the sink and the source [1-2]. Data is transmitted periodically, or on the occurrence of an event.

In this paper, we are discovering the energy aware path between the sink and the source using the parameters as used in the agricultural application. Scarcity of water is the global problem faced all over the world. Aware ness of water consumption if provided to farmers could result in conservation of water in the waterbed and avoid salinization of the soil. Adequate amount of water supply could help in good productivity of the crops. The sensors needed to detect the water contents are placed at a distance of 15cms below the ground surface. The battery utilized by the sensor has to run for a full cropping season. Energy aware routing is essential so that battery need not be replaced quite often. With this approach using ns2 simulator, energy aware path is established between the sink and the source.

## 2. RELATED WORK

Efficient path finding is essential for a longer lifetime for the batteries in the sensor network. Some work has been carried out in this area. Ref. [3] deals with the deployment of sensors to measure water content. In Ref. [4], the usage of wireless sensor network, in agriculture and cattle breeding is examined using solar power. Sensors work using either solar power, battery or grid power. As there is a likely hood of the solar panel being obstructed by the leaves, battery is the next best option for use in the agricultural field when grid power is unavailable. In the current paper we are looking into an instance where we are using battery power for the sensor to function. Ref. [5] deals with a scenario where sensor nodes are deployed in the form of a grid and readings are measured using this topology. Ref. [2], explains that a route has to be selected so that there is maximum available power, minimum energy consumption along the route. Ref. [2], discusses many routing protocols where design trade-offs between energy and communication overhead is dealt with. COMMON Sense Network is a project associated with monitoring the regulation of water supply to the field [6-8]. Based on this on- going project the path finding and path establishment algorithm NEWAODV is devised to have energy aware path to enhance the lifetime of the network using ns2.34. AODV [9-10] is used in the COMMON Sense Network. Improvisation of the protocol is carried out in this paper. Section 3 deals with the best topology to deploy sensors to enhance the network life. Section 4 deals with the criteria for simulation parameter selection. Section 5 deals with a new routing algorithm which enhances the lifetime of the network by using maximal residual energy path. In Section 6 the simulation parameter are chosen based on the design from Section 4 and used to simulate the network. Results computed are compared with the existing algorithm and it is found to be better than the existing AODV algorithm.

## 3. SELECTION OF TOPOLOGY

The mote chosen should provide long distance transmission with high sensitivity which necessitates only few motes. Simulation was carried out for various topologies [11]. The grid (square) structure though offered a marginally higher life shelf; it was at the cost of extra nodes covering the same area of interest. Physically the nodes are slightly offset from their actual strict topology. Fig. 1 shows the grid and hexagonal topology with a slight shift in their location of deployment. The topology is so chosen that they are placed in the hearing range of each other, and at the maximum only three sensors are able to listen to each other in hexagonal and only four are in the hearing range in the case of grid topology. Diagonally they are not able to listen to each other. In this paper square topology is used for analysis.

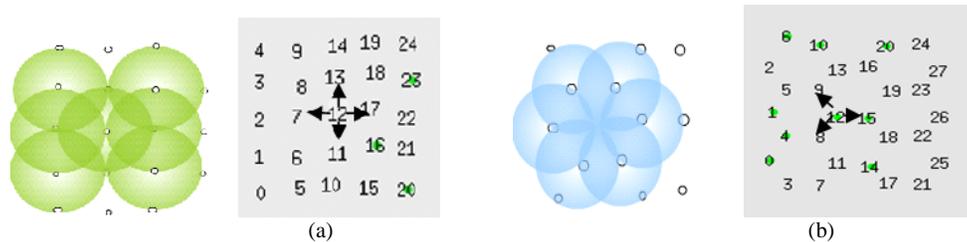

(a) (b)
Figure 1. Various topologies a) Square b) Hexagonal

## 4. PARAMETER SELECTION FOR THE NETWORK

The parameters for the simulation are as per the on-going COMMON Sense project [6]. The design is based on the simulation environment for ZigBee [12]. The sensor mote chosen is Tiny node from manufacturers of Shockfish [13]. This mote has the larger transmission range as compared to Mica motes. Following are the steps carried out in order to design the network using network simulator ns2.34.

### 4.1. Antenna Height

Tiny node supports frequency of operation of 868 and 915MHz.

$$\text{SpeedOfLight (c)} = \text{FrequencyOfOperation (f)} * \text{wavelength} \qquad (1)$$

With 868MHz. operating frequency substituting in equation (1) we have,

$3*10^8 = (868*10^6)*$

$\lambda = 0.345622$           (2)

For quarter wave antenna, antenna height is = $\lambda/4$

Substituting the value of $\lambda$ from equation (2) we get,

Antenna Height = $\lambda/4$ = 0.345622/4= 0.0864m       (3)

This is the minimum height which is essential for the proper functioning of the network. For ns simulations, the Z coordinate would give the antenna height over the ground with, X = 0, Y = 0, Z = 0.0864m. For experimental purpose the height chosen is 1m.

### 4.2 Receiver Threshold

Table1 shows the datasheet for Tiny node [13]. Based on this, the transmit power is chosen to be 5dBm. The receiver sensitivity is -104dBm @ 76.8dBm for 5dBm transmit power and range of transmission as 200m for the antenna height of 1m.

Table 1. Tiny Node Data Sheet.

| Parameter | Value |
|---|---|
| Operating Frequency | |
|   868 MHz version | 868-870 MHz |
|   915 MHz version | 902-928 MHz |
| RF Output Power | 0 to + 12 dBm |
| Data Rate | 1.2 – 152.3 kbps |
| Receiver Sensivity | |
|   @ 1.2 kbps | -121 dBm |
|   @ 76.8 kbps | -104 dBm |
|   @ 152.3kbps | -101 dBm |
| Range @ 76.8 kbps | |
|   Outdoor (1m elevation) | 200 m (+5 dBm) |
|   Indoor | 40 m (+5 dBm) |
| Current Consumption | |
|   Transmit @ +5 dBm | 33 mA |
|   Receive | 14 mA |
|   Sleep | < 1 uA |

### 4.3. Capture and Carrier Sense threshold

The least carrier sensing threshold should be at least equal to the receiver threshold value. Capture threshold is maintained at 10. This captures only the signal which is higher than other signal by 10dBm, when two signals reach the same node at same time.

### 4.4. Range of communication

According to the Two-Ray Ground Propagation model [16], the received power at a distance of d from the transmitter is given by,

$$Pr(d) = P_t G_r G_t h_r^2 h_t^2 / d^4 L \qquad (4)$$

where ht: transmitter antenna height, hr: receiver antenna height, Pr: power required to receive the signal at distance d, Pt: transmitted signal power, Gt: transmitter gain, Gr: receiver gain, d: distance from the transmitter, L: path loss.

With Pr = −104dBm =3.98e-014 W, Pt=5dBm=0.0032W, Gt=1.0, Gr=1.0, L=1.0, ht=1.0, hr=1.0m, substituting in equation (4) we have,

$3.98e\text{-}014\ W = (0.0032*1.0*1.0*(1.0)^{2}*(1.0)^{2}) / (d^4 *1.0)$

d=531m.           (5)

Therefore using a carrier sensitivity of -104dBm is equivalent to using a sensing range of around 531m around the node. As per the data specification provided in the Tiny node data sheet transmission is possible for 200m.

### 4.4. Power and Energy parameters for Tiny node

The power consumption and the energy expended in the transmission and reception of control and data signal is calculated as follows:

Two alkaline batteries of 1.5V each are used in the Tiny node. Battery voltage is,

1.5V* 2= 3V  (6)

Power consumption (W) = Current (A) * Voltage (V)  (7)

From table 1, power to transmit is 33mA, receive 14mA, sleep power is 1µA and from equation (6) and (7) we have,

The power to transmit is 33mA*$10^{-3}$ * 3V =0.099W  (8)

Power to receive is 14mA *$10^{-3}$ * 3V= 0.042W  (9)

Idle power is 2mA *$10^{-3}$ * 3V= 0.006 W  (10)

Sleep power 1µA*$10^{-6}$ * 3V = 0.000003 W  (11)

If the packet contains data of size 6bytes, containing the position of the node (x and y coordinates of 2 bytes each ) and the sensor reading (2 bytes).The data packet size at the MAC includes the common header (10 bytes), the ip header (10 bytes), data (6bytes )and the MAC header (58 bytes).

Packet size at MAC layer =10+10+6+58 bytes    =84 bytes  (12)

Transmit time (seconds) = Data Packet Size (bytes) * 8/ Data rate (bits per second)  (13)

From table 1, we have data rate = 76.8Kbps.

From equation (12) and (13) with data rate at 76.8Kbps we get,

Transmit time = 84 *8/76.8*$10^3$= 0.00875s  (14)

Energy consumed is given by,

Energy (in J) = Power (in W)*time (in secs)  (15)

Using the value from the equation (14) and equation (8) in equation (15) we get,

Energy for transmission is = 0.099W * 0.00875s

= 0.000866Ws  (16)

Using the value from equation (14) and (9) we get,

Energy for reception is = 0.042W *0.000848s

= 0.000367Ws  (17)

During a single data transmission and reception the amount of energy expended in a node is (0.000866Ws+ 0.000367Ws) = 0.001233Ws.

Fig. 2 shows the battery discharge characteristics for alkaline battery [14].

From table 1, power to transmit is 33mA, receive 14mA. So, on an average for one transmission and one reception,

Average current is (33+14) mA /2 =23.5mA.  (18)

The number of hours of operation is 80 hours for a current consumption of 23.5mA from the discharge characteristics as shown in fig.2.

Substituting the value from equation (18) in (7) for a 3 volt battery supply we get,

23.5*$10^{-3}$ *3V=0.0705W  (19)

Substituting in equation (19) into equation (15) the available energy for 80 hours of operation is

Energy (in J) = Power (in W)*time (in secs)
$$= 0.0705W*(80*60*60)$$
$$= 20304 \text{ Joules or } 20304Ws \qquad (20)$$

This is the initial energy available in all the nodes in the network.

The simulation is carried with the different values of initial energy in each node.

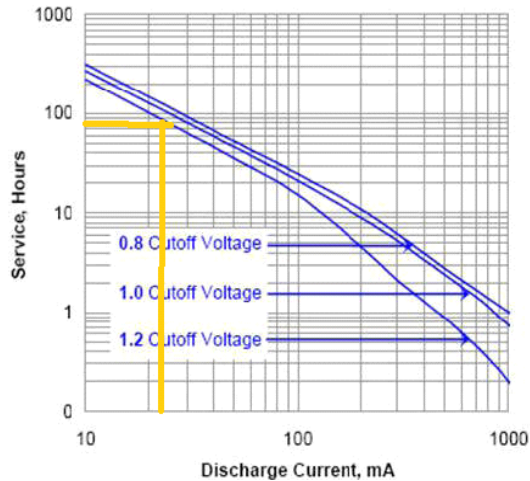

Figure 2. Discharge characteristics [14]

## 5. ROUTING ALGORITHM

The ongoing project COMMON-Sense uses AODV protocol. As the nodes are deployed in a static manner, the frequent broadcast and checking for the validity of neighbors is unnecessary [14]. The route however follows the minimum hop count but also based on available energy in the neighbors.

Fig. 3 shows a square topology of 9 nodes. The nodes are provided with increasing order of energy from bottom to top and then from left to right. Node 7 has the least energy and node 5 the largest amount of energy. With 7 as the source and 5 is the destination, minimum hop count between source and destination 4 and number of path available is 6.

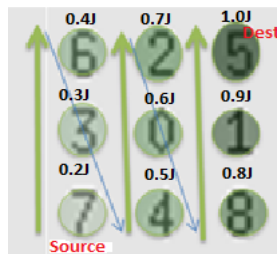

Figure 3 Topology with 9 nodes

Among the various paths, the path which has maximum total energy residual energy along its path in the order of increasing residual energy is as shown in the table 2. Based on the above distribution of energy, the path which has least residual energy along its path is from 7→3→6→2→5. The highest residual energy is along 7→4→8→1→5.

Table 2. Residual energy distribution

| Paths traversed | Total residual energy along the path |
|---|---|
| 7→3→6→2→5 | 0.2+0.3+0.4+0.7+1.0=2.6J |
| 7→3→0→2→5 | 0.2+0.3+0.6+0.7+1.0=2.8J |
| 7→3→0→1→5 | 0.2+0.3+0.4+0.9+1.0=2.8J |
| 7→4→0→2→5 | 0.2+0.5+0.6+0.7+1.0=3.0J |
| 7→4→0→1→5 | 0.2+0.5+0.6+0.9+1.0=3.2J |
| 7→4→8→1→5 | 0.2+0.5+0.8+0.9+1.0=3.4J |

The NEWAODV algorithm chooses the path, where the residual energy is maximum along its path. Considering the topology as shown in fig. 3, the number of minimum hop path from node 7 to 5 is 6. This implies there could be 6 control signal search path reaching the destination node 5 and the path which has maximum total energy along the path has to be chosen.

With AODV all the control signals do not reach the destination. Control signal whose hop count matches with the already existing path is dropped. In the current algorithm, control signals are forwarded even though the hop count are matching if the total energy along the path is more than the previous found total energy path.

The following pseudo code shows the algorithm used to search for energy aware path.

1. Start the path search using the control signal with a new sequence number from the source to the destination.
2. Initialize the total energy in the path to zero.
3. In each of the traversed node,

    3.1 Make an entry or update the entry in the routing table, for each source destination combination, for each of the node traversed provided,
    3.1.1 The route to be searched is a new one with a new sequence number then,
    3.1.2 With minimum hop count from those found so far between the corresponding source destination combination then,(go to step 3.1.3)
    3.1.3 If the total energy computed now is larger than the already computed total energy which is stored in the routing table, then, (go to step 3.1.4)
    3.1.4 If the above three conditions are satisfied, update the sequence if it is a new one, update the hop count if it is minimal, update the total energy value if it maximum along the path with minimum hop count, then, (go to step 3.1.5)
    3.1.5 For each entry made in the routing table a reverse routing table entry is made for the node from where the signal is obtained.
    3.1.6 Else go to step 3.2
    3.2 Otherwise drop the control packet.
4. This process is repeated until it reaches the destination to drop the control packet. If the hop count is more than the time to live (the maximum hop count possible for the network), then drop the packet as there is no route from this node to the destination.

    Once the packet reaches the destination another control packet is started in the reverse direction from the destination to the source

5. The node traverses the path as specified in the reverse routing table to reach the destination.
6. If multiple paths exist up to the source from the destination, the most efficient one with maximum residual reverse path is chosen.

    Once the path with maximal total energy for path establishment is chosen,

7. Data transmission starts from the source to the destination

8. In each node along the path,
8.1 Forward the data until it reaches the destination from the source.
9. This process is repeated until there is a node or path failure upon which a new path search is invoked with total energy initialized to 0.
10. The process of searching for a new path continues until there are no more paths available from the source to the destination.

## 6. SIMULATION AND RESULTS

The simulation is carried out using ns2.34. Following is the scenario observed during path search. The source is node 23 and the sink is node 21. With initial energy of 5J and running for an 80 hour period and time interval of 300sec., only the first path between the source and destination is alive for long interval of time. Fig. 4a shows a path with 415 transmissions. Subsequent to that path as in fig. 4b, 4c, 4d, 4e, 4g had 1 transmission of data and 4f had 2 transmissions. Almost all the paths are having 1-2 rounds of data transmission. This consumes large amount of energy for path search with each round, and data is sent for only one round. Yellow colored node represents the nodes which have currently lost their energy. The orange colored nodes show all nodes which have lost energy and the network has partitioned with no route between the source and the destination.

Using the above algorithm, if the residual energy along the path is large, the network is likely to transfer data for a longer time along a selected path. The number of new path search is decreased conserving enough energy in the network due to path search. Another advantage is the formation of disjoint network can be postponed due to data transmission in higher residual energy path.

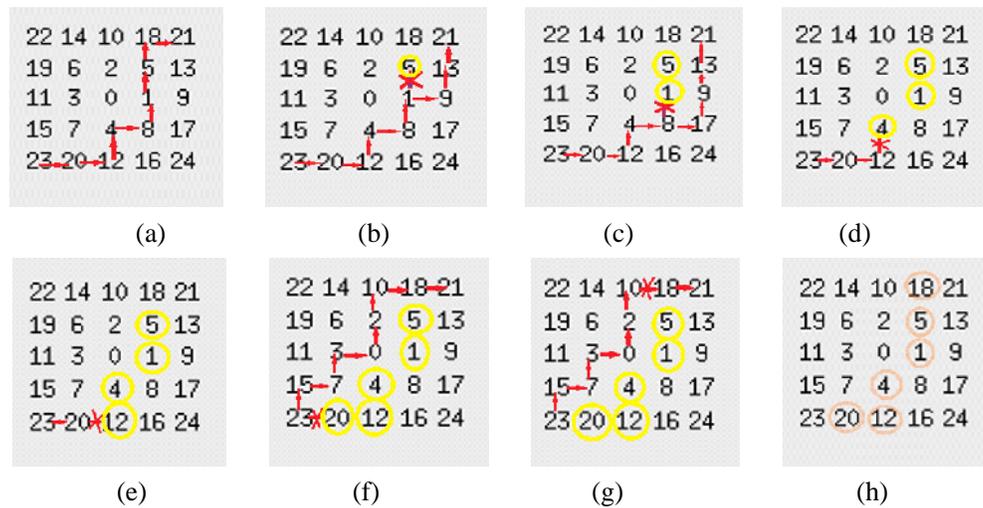

Figure 4. Sequence of various paths for sending the data

For the distribution of energy as shown in Fig. 3 for a 5*5 topology the path chosen by the network is as shown in fig. 5. Among the various paths to the destination node 21, from source node 23, the path with maximal residual energy is 23→20→16→24→17→9→13→21 as in fig 5a. Node 23 being source is assigned a higher value of energy as compared to its neighbors, otherwise node 23 dies first. Among 20 and 15, 20 has higher energy. Among 7 and 12, 12 has higher and so on with this concept the first path is maintained for long interval of time. Along this path as the node 20 is having lowest energy in this path, this node loses energy first and a new path 23→15→7→4→8→1→5→13→21 is chosen as shown in fig. 5b. While choosing node 13 and 18, node 13 has higher residual energy than 18 and hence path is established through node 13. As the nodes along the first path have lost considerable amount of energy due to sending of data none of the nodes except 13 is chosen in the new path search. The next node to lose energy is 15 and this has resulted in no path from the source to the sink. If we look at the existing AODV protocol, the path chosen is as shown in fig. 5c followed by fig. 5d. The corners nodes are not normally chosen in AODV as it normally encounters collision. To avoid this, the delay of 0.05s is introduced in our routing algorithm between two successive transmissions of control signal instead of the normal 0.01s. From the fig c and fig d it is clear that the two paths chosen by the AODV are less energy path as compared to paths as in fig a and fig b corresponding to NEWAODV algorithm. Hence more number of transmissions is possible in NEWAODV as compared to AODV.

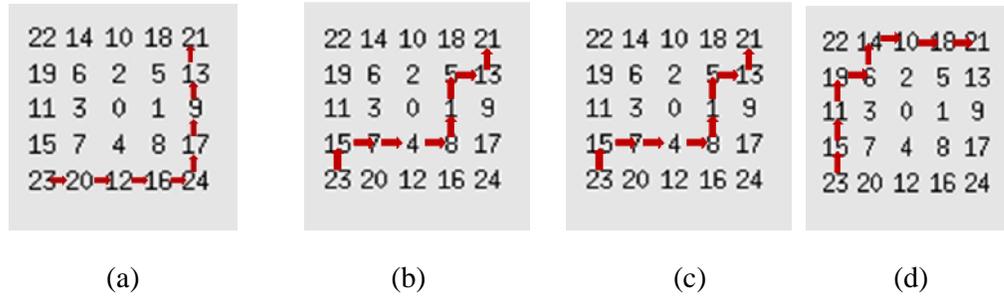

(a) (b) (c) (d)

Figure 5. Paths selected by the new algorithm

Table 3. Simulation parameters

| Radio Parameters | | Simulation parameters | |
|---|---|---|---|
| Radio frequency | 868 MHz | Date reading interval | 300secs. |
| Antenna Height | 1m (min. reqd. height 0.0819m) | Nodes | 9/25 |
| Antenna Type | Omnidirectional – Quarter wave | Topology | Square/Hexagonal |
| Transmit Power | 3.16 mW =5dBm | MAC | 802.11/802.15.4 |
| Receive Power | −104dBm@ 5dBm=3.98e-14W | Queue | Drop Tail |
| Carrier Sense Threshold | −104dBm@ 5dBm | Queue size | 150 |
| Capture Threshold | 10 dBm | Protocol | AODV/ NEWAODV |
| Gain of transmitting/Receiving antenna | 1 | | |
| | | | |
| **Sensor parameters (Tiny Node)** | | **Battery (Alkaline battery of 1.5V)** | |
| Receive Power | 0.099W=19.95dBm | Battery supply | 3V with 2 AA sized battery |
| Transmit Power | 0.042W=16.23dBm | Power consumption | 0.0705W for 23.5mA discharge current |
| Sleep Power | 0.000003W= -25.2 dBm | Energy consumption | 20304J for 80 hrs. of operation |
| Idle Power | 0.006W =7.78dBm | | |

Table 4. Simulation results

|  | AODV | | NEWAODV | |
|---|---|---|---|---|
| **Case I** | **Topology :**Square with 9 nodes (fig.3) with source at node 0 and sink at node 9<br>**Initial minimum energy in each node :** 0.2J<br>**Difference in energy between nodes:** 0.1J | | | |
| Delivery ratio | 99.4% | 349/351 | 99.7% | 350/351 |
| Average Energy consumed | 0.2330 J | 0.1749 J –I<br>0.0332 J –T<br>0.0247 J –R | 0.2328 J | 0.1745 J – I<br>0.0333 J – T<br>0.0248 J – R |
| **Case II** | **Topology:** Square with 9 nodes (fig.3)<br>**Initial minimum energy in each node:**0.3 J<br>**Difference in energy between nodes:** 0.08J | | | |
| Delivery ratio | 98.86% | 348/351 | 99.4% | 349/351 |
| Average Energy consumed | 0.3825 J | 0.3135 J –I<br>0.0313 J –T<br>0.0353 J –R | 0.3826 J | 0.3135 J – I<br>0.0335 J – T<br>0.0335 J – R |
| **Case III** | **Topology:** Square with 25 nodes (fig 5)<br>**Initial minimum energy in each node :**0.02 J<br>**Difference in energy between nodes:**0.01J | | | |
| Delivery ratio | 34.5% | 74/214 | 35.2% | 75/213 |
| Average Energy consumed | 0.07478 J | 0.0619 J –I<br>0.0059 J –T<br>0.0068 J –R | 0.074013 J | 0.0613 J – I<br>0.0061 J – T<br>0.0064 J – R |
| Case I, II,III for a simulation period of 105000 seconds using Mac 802.11 protocol with source at 23 and sink node 21 for case II and III | | | | |
| Case IV,V for simulation period of 10500 seconds using Mac 802.15.4 protocol with source 19 and sink node 17 ( beacon enabled ) square topology with 25 nodes with pan coordinator node 0 | | | | |
| **Case IV** | **Beacon enabled coordinator nodes:** 5,6,7,8,9,10,11,12 (fig 5)<br>**Initial minimum energy in each node:**0.2 J<br>**Difference in energy between nodes:**0.2 J | | | |
| Delivery ratio | 94.1% | 32/34 | 97.0% | 33/34 |
| Average Energy consumed | 1.2743 J | 0.0000 J –I<br>0.4850 J –T<br>0.7891 J –R | 1.2544 J | 0.0000 J – I<br>0.4845 J – T<br>0.7699 J – R |
| **Case V** | **Beacon enabled coordinator nodes:**5,6,7,8,19,13,17,15 (fig 5)<br>**Initial minimum energy in each node:**0.2 J<br>**Difference in energy between nodes:**0.06 J | | | |
| Delivery ratio | 88.23% | 30/34 | 94.12% | 32/34 |
| Average Energy consumed | 0.5015 J | 0.0000 J –I<br>0.2176 J –T<br>0.2839 J –R | 0.5075 J | 0.0000 J – I<br>0.2172 J – T<br>0.2902 J – R |

Table 3 shows the simulation parameters as obtained from equations (3), (5), (8), (9), (10), (11) ans (20). These values are used for the computation of the results. The data value is monitored for every 5 minutes= 300seconds.

Table 4 shows the simulation results with the average energy consumption representation as I-Idle, T-Transmit, R- Receive

From the table 4 we find that whether we used 802.11 or 802.15.4 at the MAC layer, the performance of the NEWAODV algorithm is an improvement on the existing AODV algorithm. The delivery ratio has improved along with the average reduction in the energy consumption. This is due to the reduced number of path searches. There could also be certain nodes in the path which have low energy, participated in establishing the path, but do not have sufficient energy to transmit data. These paths which have been established serve no purpose.

The disadvantage of this algorithm as compared to AODV is, in AODV, higher or same hop count paths packets are dropped. But with this approach large amount of energy is required for path search. But this gets compensated with more number of transmissions in higher residual energy path.

## 7. CONCLUSIONS

Sensor nodes are deployed in the agricultural field to measure the water content in the soil. To avoid excess usage of water in drought hit areas, the amount of water that is supplied to the field should be regulated. The sensor should be sensing the data value once in every 5 minutes and this is done for one cropping season lasting around 6 months. This avoids replacing the battery often and avoids fresh new path from being discovered quite often. This requires that the network be alive for longer interval of time. In order to accomplish this, path has to be so chosen that, it should try to conserve as much energy as possible. This could be possible if we could reduce the number of unnecessary transmissions and receptions. This could also imply that frequent path searches should be avoided. This necessitates that the path once established should be utilised for a longer interval of time. A path could be sustained for longer interval, if it is a path which has higher energy along the path when data is sent. The new algorithm tries to find a path so that the energy along its path is maximum, by checking the residual energy in the node when new path is established. The distance between the nodes are maintained such that, it does not receive unnecessary reception from many neighbours. At most there could be 4 neighbours contributing to reception of broadcast signal. The distance between the base station and sensing nodes are large, multihop communication is essential and hence minimum number of hop count is also taken into consideration when a new path is established. Results show that choosing a maximal residual energy path can increase the number of times data can be transmitted along a chosen path which also avoids new path search often. It is also observed from the results that the average amount of energy consumed in the network is less as compared to the existing AODV protocol. When the network is likely to be partitioned into disjoint network, it is desirable to have as many data transmissions as possible. This is due to the reason that though there could be some energy in the network during partitioning it may not contribute to the data transmission and could become a waste. It is observed that more energy is consumed during idle state. Adaptive sleep can be introduced to avoid consumption of energy during idle state.

**Authors**

Smitha N. Pai is an Assistant Professor in the Department of CSE at MIT Manipal. She obtained her M.Tech. in CSE from Manipal University in 2002 and B.E in E&C from Mysore University. She is currently pursuing her PhD. from Manipal University. Her main interest lies in wireless sensor, adhoc networks.

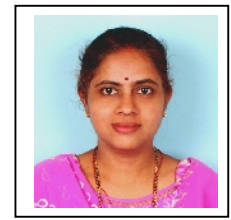

Dr.K.Chandrashekar Shet obtained his B.E, M.Sc.(Engg) and Ph.D degrees from Mysore University, Sambalpur University and IIT Bombay in the years 1972, 1979 and 1987 respectively. He is working in NITK Surthakal since 1980 and presently he is the professor in the Dept.of Computer Sc.& Engg, and Dean (Faculty welfare). He has published around 250 papers, in National, International, journals/conferences. Besides, he has published three books, on Micro-processors, Software Engg. & quality Assurance. He has produced 10 Ph.D professionals and currently six are pursuing research under his guidance leading to Ph.D.

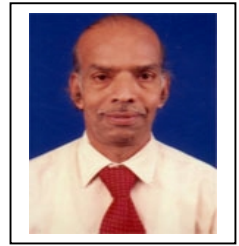

**H. S. Mruthyunjaya** has completed his bachelor degree in Electronics and Communication Engineering from Mysore University in 1988 and obtained his masters degree in Electronics and Control Systems Engineering from Birla Institute of Technology and Science, Pilani in 1994. He has a Ph.D in Electronics and Communication Engineering conferred by Manipal University for his thesis entitled 'Performance Enhancement of Optical Communication Systems and Networks using Error Control Techniques'. He is currently serving as a Professor in the Department of Electronics and Communication Engineering, Manipal Institute of Technology, Manipal, India where he joined as a Lecturer in the year 1998. He has done research on countering non-linear effects and other noises in WDM all-optical networks by employing error control coding techniques. His areas of major interests are the Optical Fiber Communication systems, Fiber Optics, Photonic Crystal Fibers, WDM networks and systems, Electromagnetic theory & General areas of Digital Communication Systems. He has authored or co-authored over Forty three technical papers in refereed journals and International conference proceedings. He is a Fellow of the Institution of Engineers (India) and member of Indian Society for Technical Education.

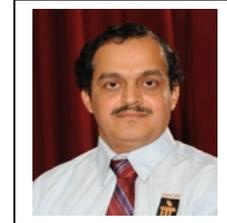